\documentclass[12pt,preprint]{aastex}

\usepackage{graphics}
\usepackage{epsfig}

\shorttitle{Tycho SN 1572: A Naked Ia SNR}
\shortauthors{Tian \& Leahy}

\begin{document}

\title{Tycho SN 1572: A Naked Ia Supernova Remnant without Associated Ambient Molecular Cloud} 
\author{W.W. Tian\altaffilmark{1,2}, D.A. Leahy\altaffilmark{2}}
\altaffiltext{1}{National Astronomical Observatories, CAS, Beijing 100012, China; email: tww@bao.ac.cn}
\altaffiltext{2}{Department of Physics \& Astronomy, University of Calgary, Calgary, Alberta T2N 1N4, Canada}
 
\begin{abstract}
The historical supernova remnant (SNR) Tycho SN 1572 originates from the explosion of a normal Type Ia supernova which is believed to have originated from a carbon-oxygen white dwarf in a binary system. We analyze the 21cm continuum, HI and $^{12}$CO-line data from the Canadian Galactic Plane Survey in the direction of SN 1572 and surrounding region. We construct HI absorption spectra to SN 1572 and three nearby compact sources. We conclude that SN 1572 has no molecular cloud interaction, which argues against previous claims that a molecular cloud is interacting with the SNR. This new result does not support a recent claim that dust, newly detected by AKARI, originates from such a SNR-cloud interaction. We suggest that the SNR has a kinematic distance of 2.5 - 3.0 kpc based on a nonlinear rotational curve model. Very-high-energy $\gamma$-ray emission from the remnant has been detected by the VERITAS telescope, so our result shows that its origin should not be an SNR-cloud interaction. Both radio and X-ray observations support that SN 1572 is an isolated Type Ia SNR.  
\end{abstract}


\section{Introduction and Data}
Recent observations have revealed that the historical Supernova Remnant (SNR) Tycho SN 1572 belongs to the class of Type Ia SN by detecting its optical spectrum near maximum brightness from the scattered-light echo \citep{Kra08}. A red subgiant has been suggested to be the possible surviving companion of the supernova in a close binary system \citep{Rui04}. However, the evolutionary path of the progenitor is still not understood, and this association has been questioned \citep{Fuh05, Iha07}. SN 1572 is a natural candidate for high energy observations. Non-thermal X-ray emission and thin filamentary structures in the remnant are believed to be associated with high-energy electron acceleration \citep{Bam05, Kat10}. \citet{War05} studied the shock dynamics using Chandra observations and suggested there is efficient hadronic particle acceleration in the remnant. Weak TeV emission coming from the direction of SN 1572 is newly detected by the VERITAS telescope \citep{Acc10}, although confirmation by other instruments (e.g. MAGIC) is required.
   
Young, massive core-collapse SNe are usually not far away from their parent molecular clouds since their progenitors evolve very quickly (a few million years). Therefore, it is expected that many Galactic Type II/Ibc SNRs are associated with large molecular clouds. Recently, \citet{Jia10} cataloged more than 60 possible SNR-cloud interaction systems. As the only known Type Ia remnant in this catalogue, SN 1572 has been proposed to be interacting with dense ambient (atomic/molecular) clouds toward its Northeast (NE) \citep{Rey99, Lee04, Cai09}. Cold dust overlapping the eastern part of SN 1572 has been detected \citet{Ish10} and taken as evidence of a possible interaction between SN 1572 and a molecular cloud. Extended TeV emission detected in several SNRs has been suggested to originate from the interaction between the SNR shock and an adjacent CO cloud \citep{Eno02, Aha04, Alb07, Tia08, Cas10, Tav10}. Is TeV emission from SN 1572 associated with such a SNR-cloud interaction?

In this letter, we take advantage of the 21cm continuum, HI and $^{12}$CO-line data from the Canada Galactic Plane Survey (CGPS) in the direction of SN 1572. We study if there exists such an interaction responsible for the TeV emission. Details on the CGPS are given in  \citet{Tay03}, the analysis methods are described in our previous papers \citep{Tia07, Tia10}.

\section{Results and Analysis}
\subsection{Continuum images and spectra}
In Figure 1, we show the 1420 MHz continuum image around SN 1572. In Figures 2 and 3, we show the HI and $^{12}$CO spectra extracted for SN 1572 and three nearby compact sources. As SN 1572 is extended, we employ our well-tested methods in order to obtain reliable HI absorption spectra  \citep{Joh09, Zho09, Lea08}. First, our background region is chosen to be near the continuum peak. This is shown in Fig. 1, the source and background spectra are shown by the solid-line and dashed-line boxes (the background area excludes the source area). This minimizes the difference in the background HI distribution along the two lines of sight. Second, we select four areas to extract the HI absorption spectra. This maximizes real absorption features in SN 1572. Third, we build HI absorption spectra for three compact sources near SN 1572. Finally as a comparison with the HI spectra we show the $^{12}$CO emission spectra. This allows one to constrain the distance to SN 1572 and understand the HI absorption spectra. 

From the HI spectra, we find that the highest absorption velocity is -53 km s$^{-1}$ in all spectra towards SN 1572 (see Fig. 2a and Fig. 3).  
 The highest absorption velocities towards the other sources are $\approx$ -110 km s$^{-1}$. These are all much higher than that of SN 1572, so they must be behind SN 1572 and are probably extragalactic. For the three compact sources, any HI emission that has brightness temperature of above 20 K shows associated HI absorption in the whole velocity range. HI emission with peak brightness temperatures of 70 to 90 K, i.e. at -68 km s$^{-1}$ for G119.72+2.4, -58 km s$^{-1}$ for G119.71+1.12 and -57 km s$^{-1}$ for G120.56+1.21, has associated deep HI absorption. This is not this case for SN 1572 (see Figs. 2a and 3). There is no associated HI absorption for the HI emission peak at -60 km s$^{-1}$, and also for the other lower  peaks at -46, -90 and -98 km s$^{-1}$. Fig. 2 also shows the CO emission spectra. For G120.56+1.21, the high brightness-temperature CO component with a peak at -60 km s$^{-1}$ has strong associated HI absorption. However, for SN 1572, the bright CO molecular cloud with a peak at -64 km s$^{-1}$ does not produce any HI absorption. 
 
\begin{figure} 
\vspace{50mm} 
\begin{picture}(80,80)
\put(-80,250){\includegraphics{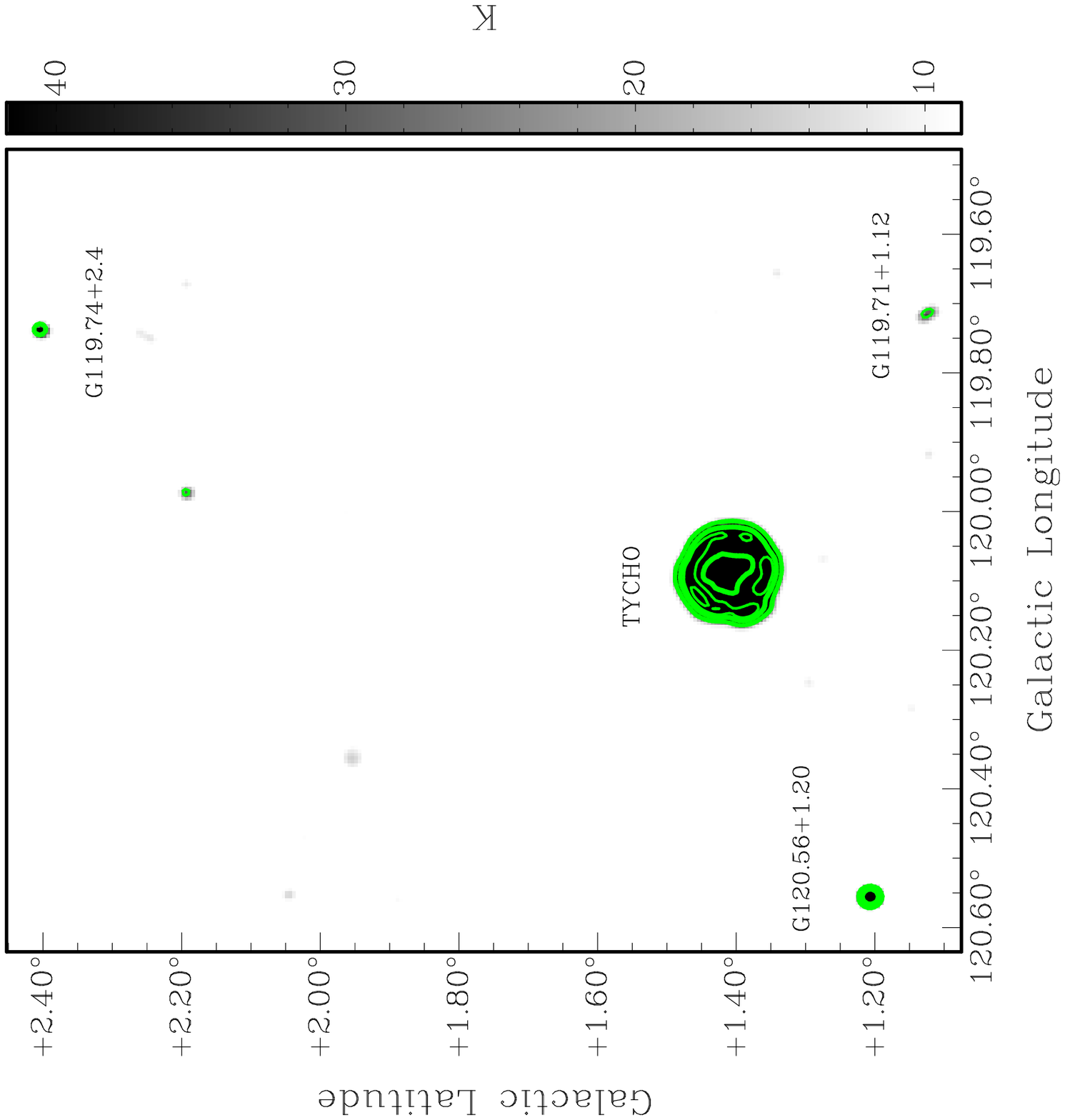}}
\put(245,0){\includegraphics{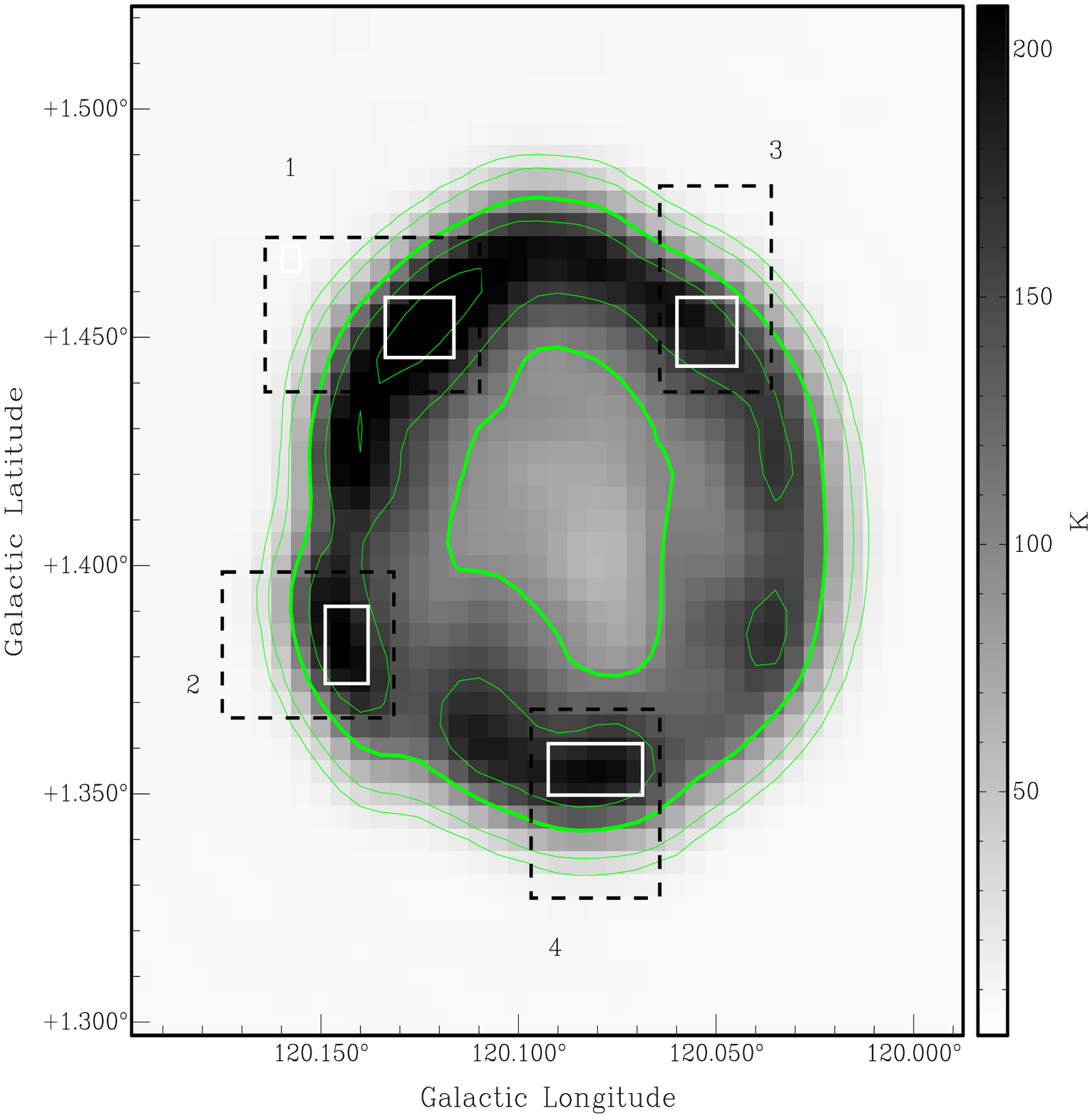}}
\end{picture}
\caption{The 1420 MHz continuum image with contours (22, 40, 100, 160, 225 K) around SN 1572 (left) and the zoomed SN 1572 image. North is up, the East is to the left}  
\end{figure} 

\begin{figure}
\vspace{185mm}
\begin{picture}(80,80)
\put(0,455){\includegraphics{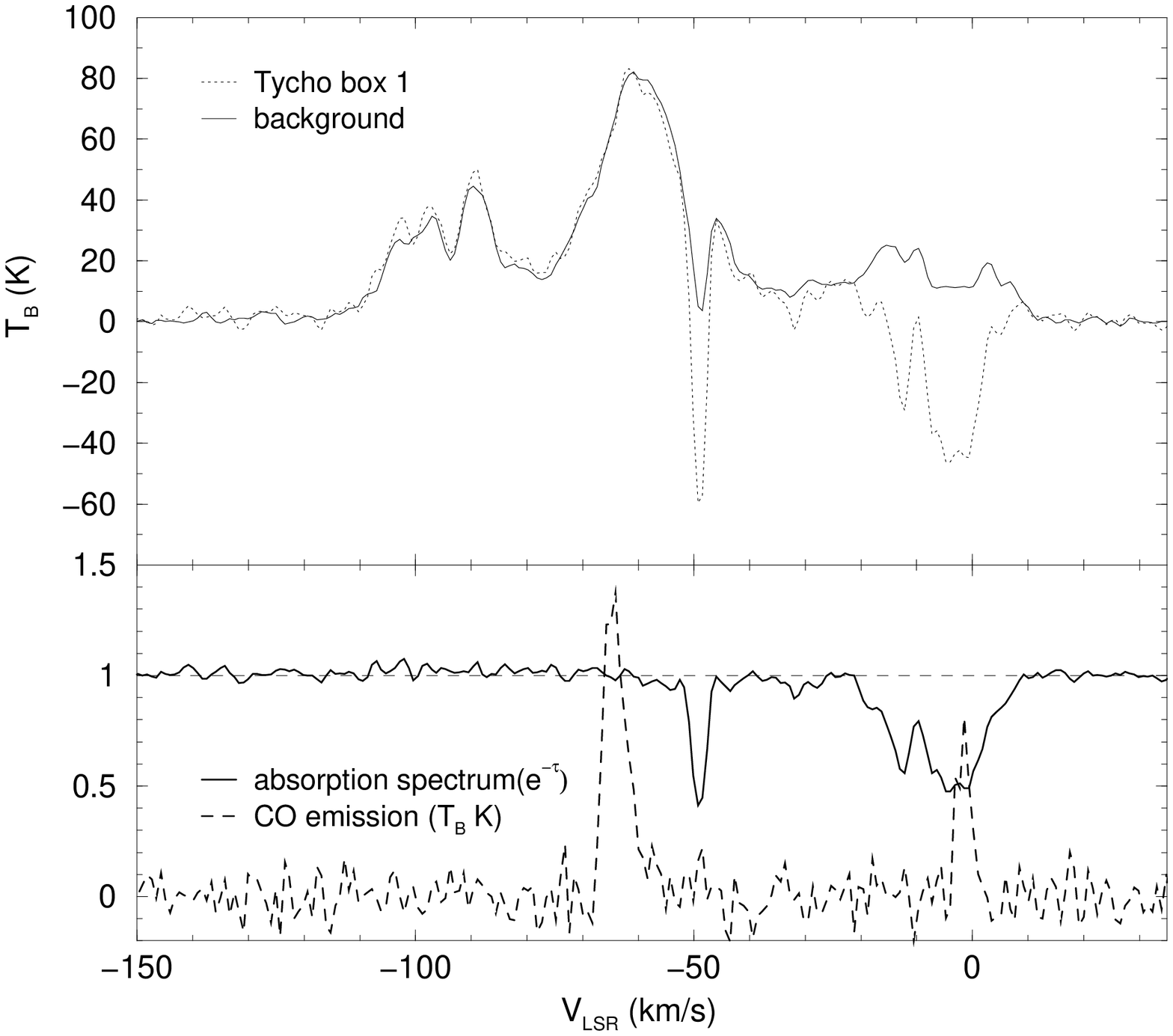}}
\put(0,300){\includegraphics{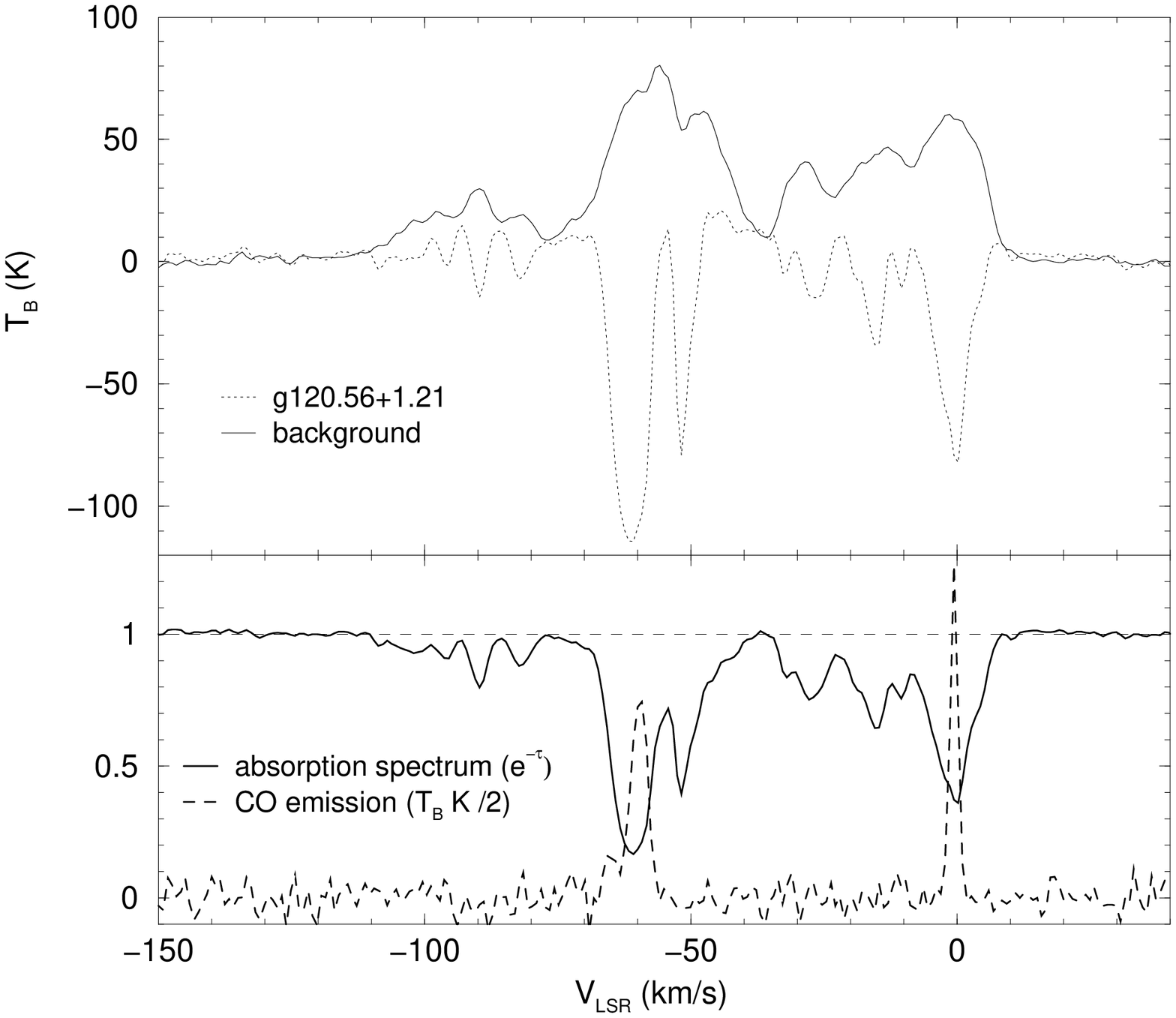}}
\put(0,145){\includegraphics{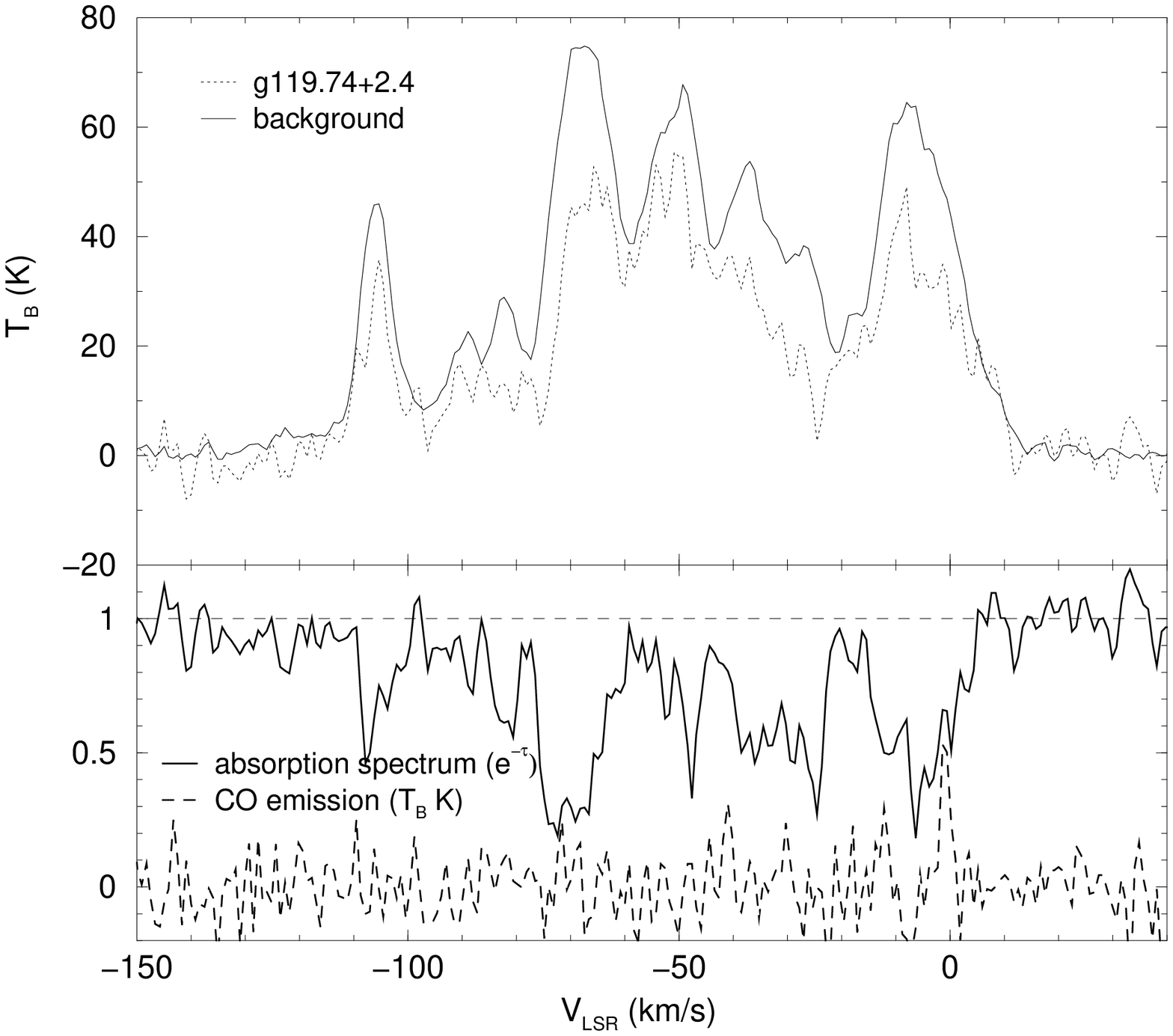}}
\put(0,-10){\includegraphics{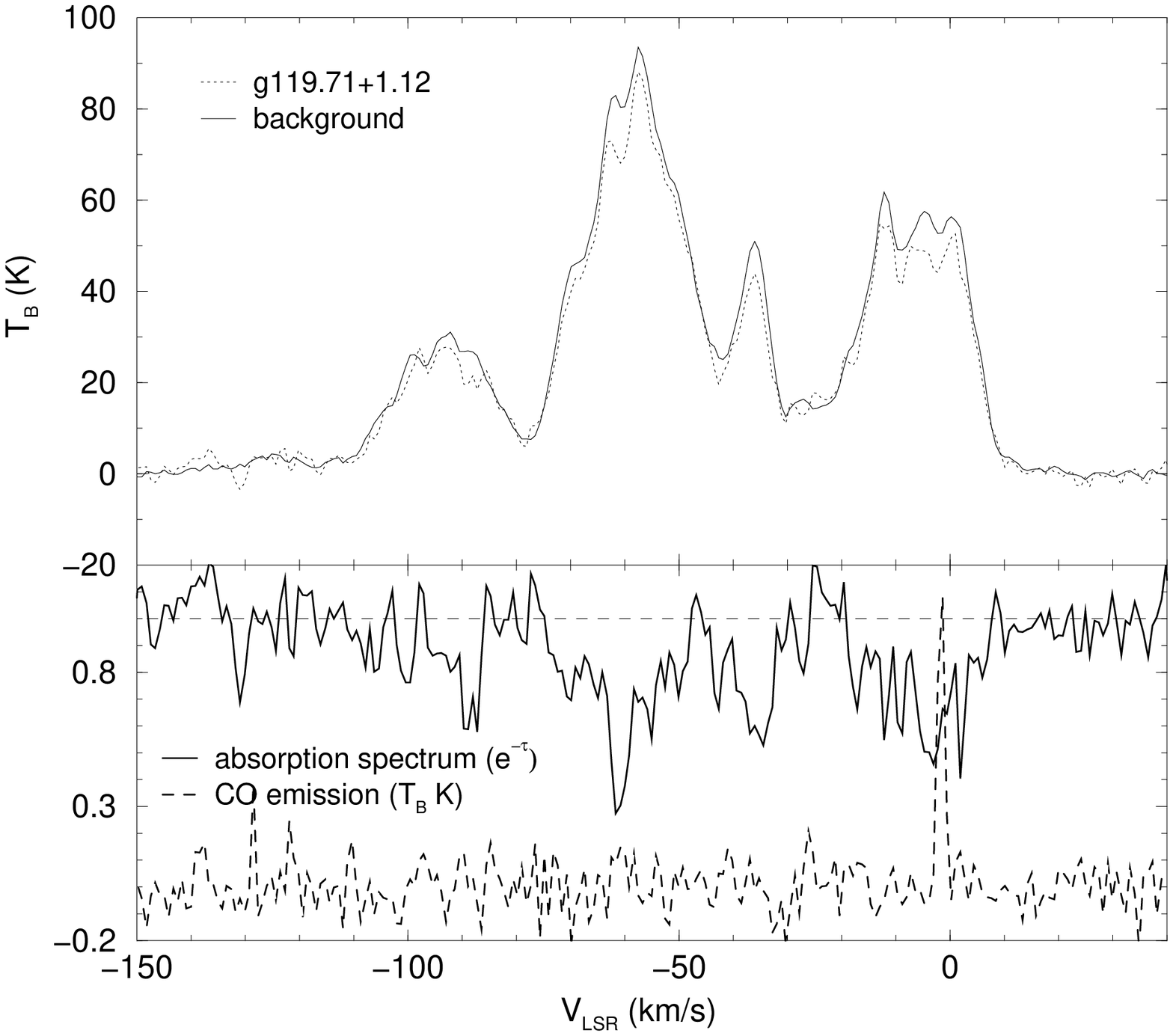}}
\end{picture}
\caption{HI emission (upper panel) and associated HI absorption and $^{12}$CO emission spectra (lower panel) of SN 1572 and compact sources: G120.56+1.20, G119.74+2.4 and G119.71+1.12}
\end{figure}

\begin{figure}
\vspace{120mm}
\begin{picture}(80,80)
\put(0,340){\includegraphics{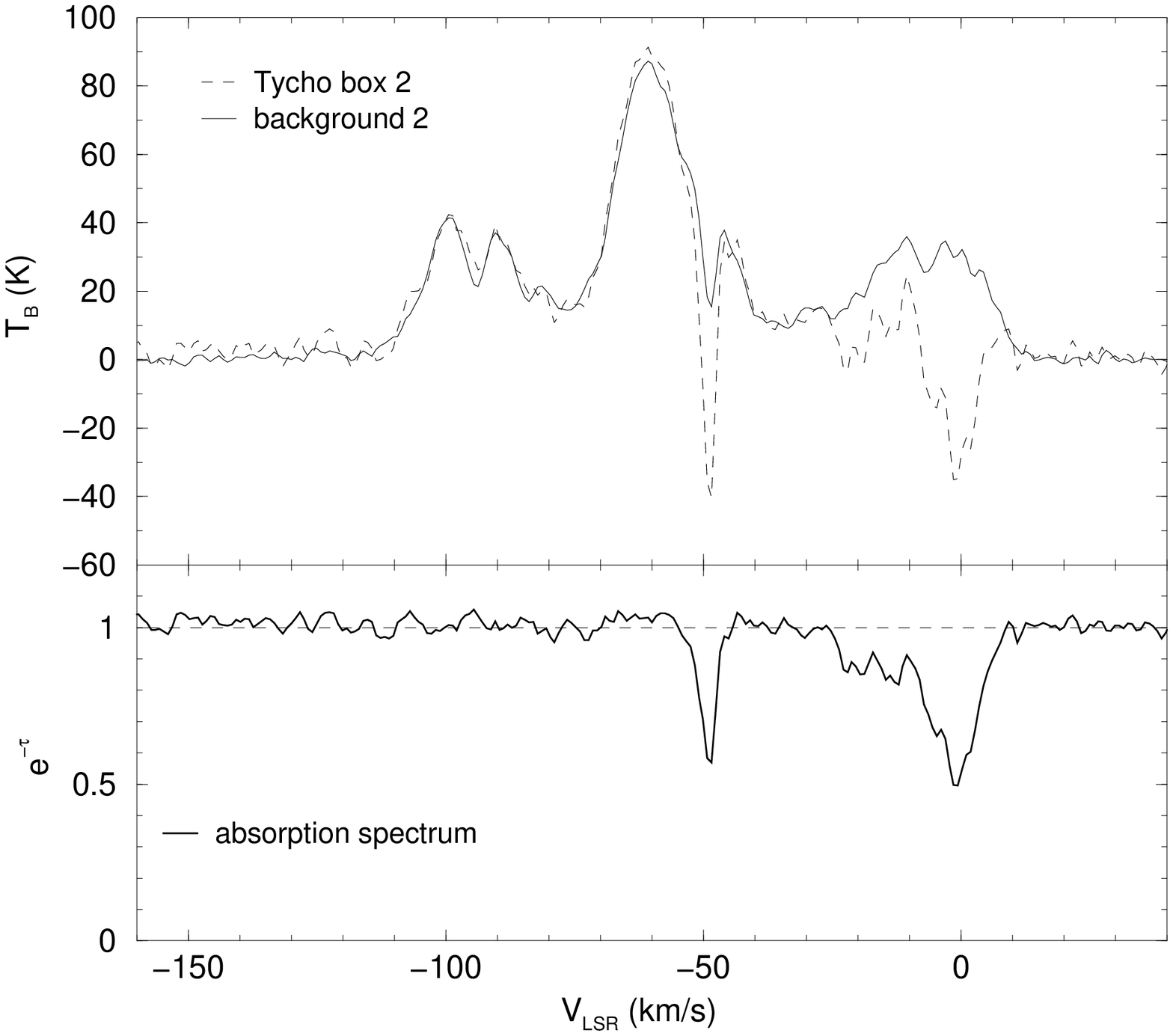}}
\put(0,160){\includegraphics{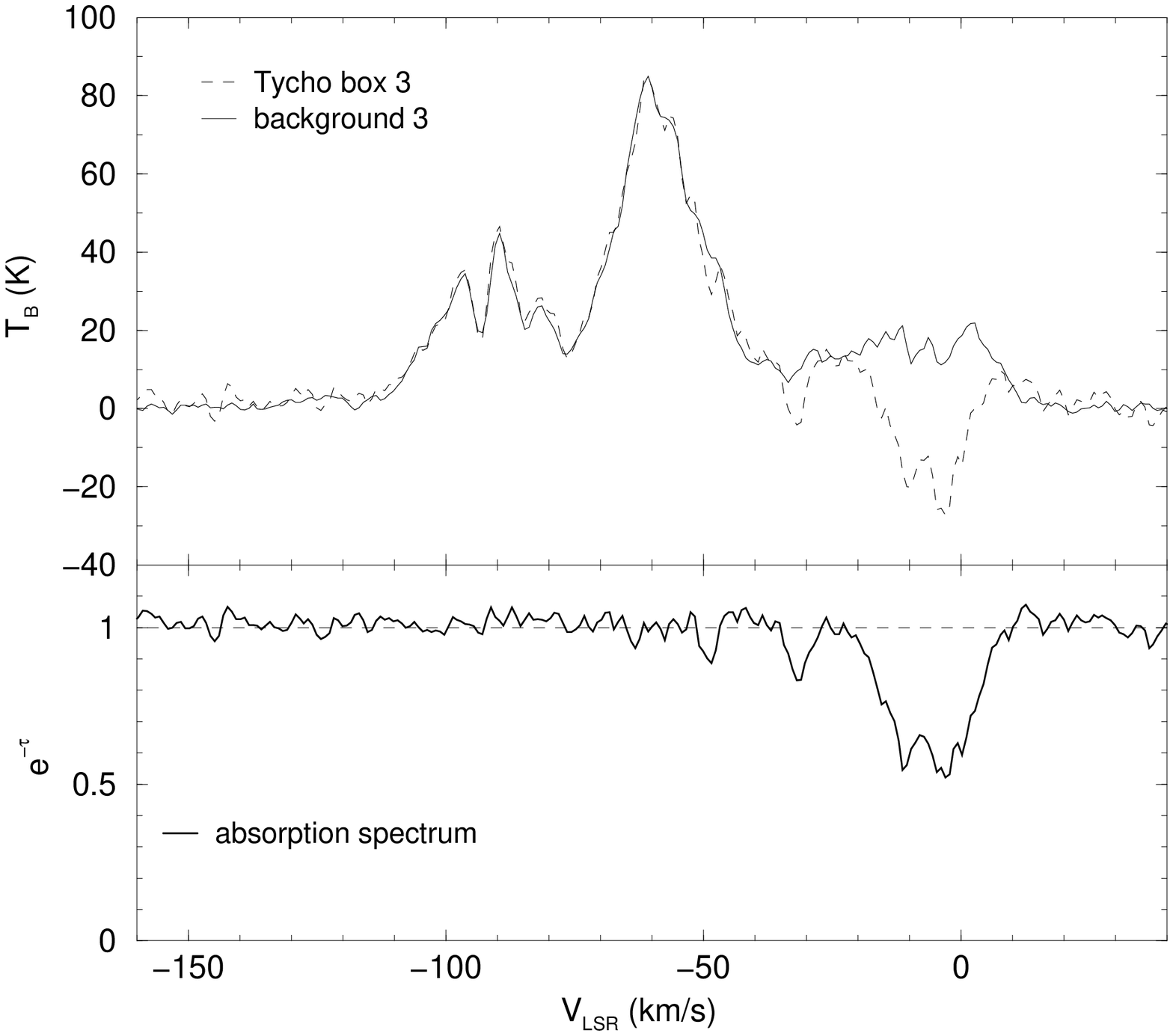}}
\put(0,-20){\includegraphics{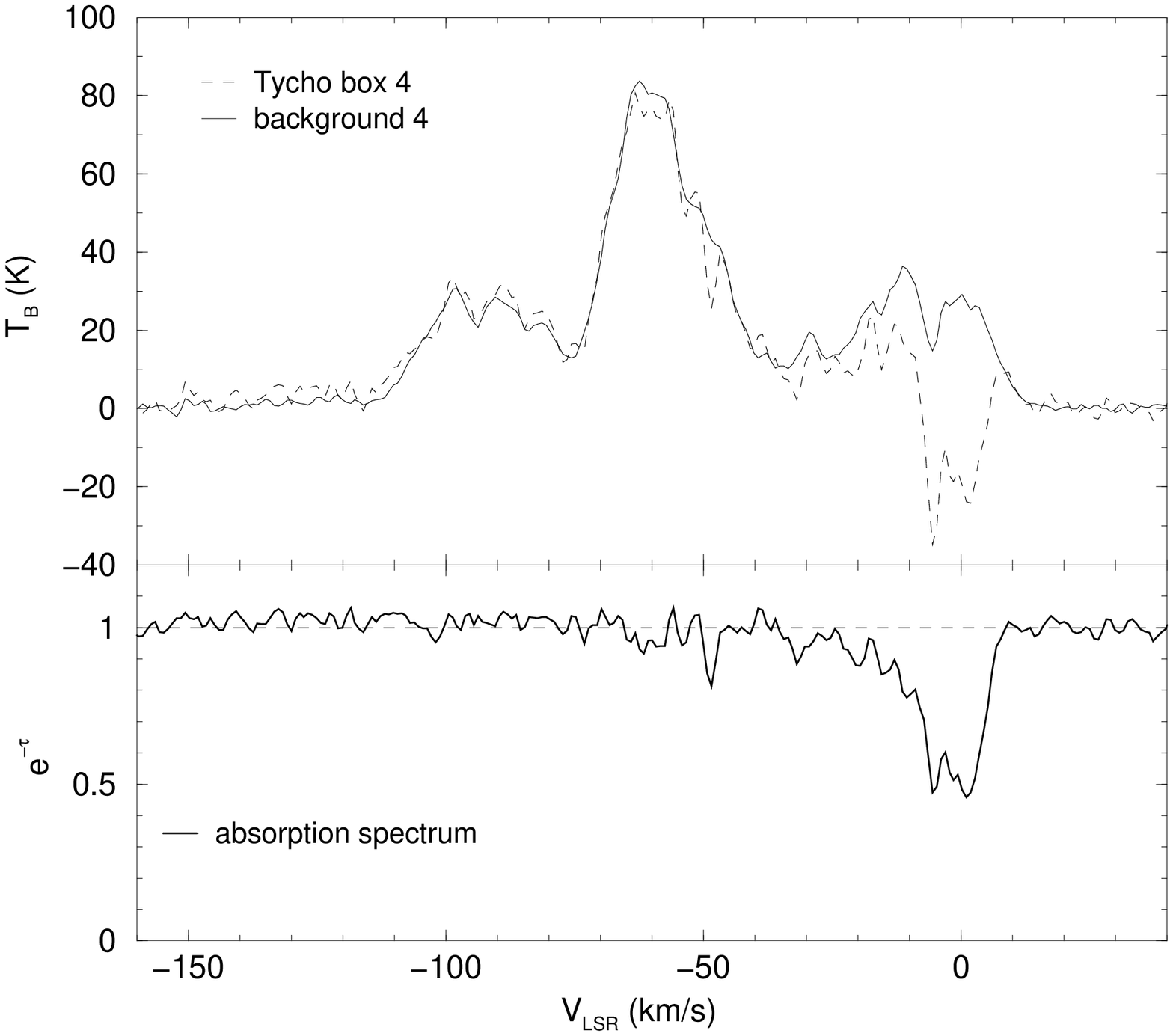}}
\end{picture}
\caption{HI spectra extracted from SN 1572 boxes 2, 3 and 4}
\end{figure}

\subsection{HI and CO channel maps}

\begin{figure}
\vspace{180mm}
\begin{picture}(80,80)
\put(-40,630){\includegraphics{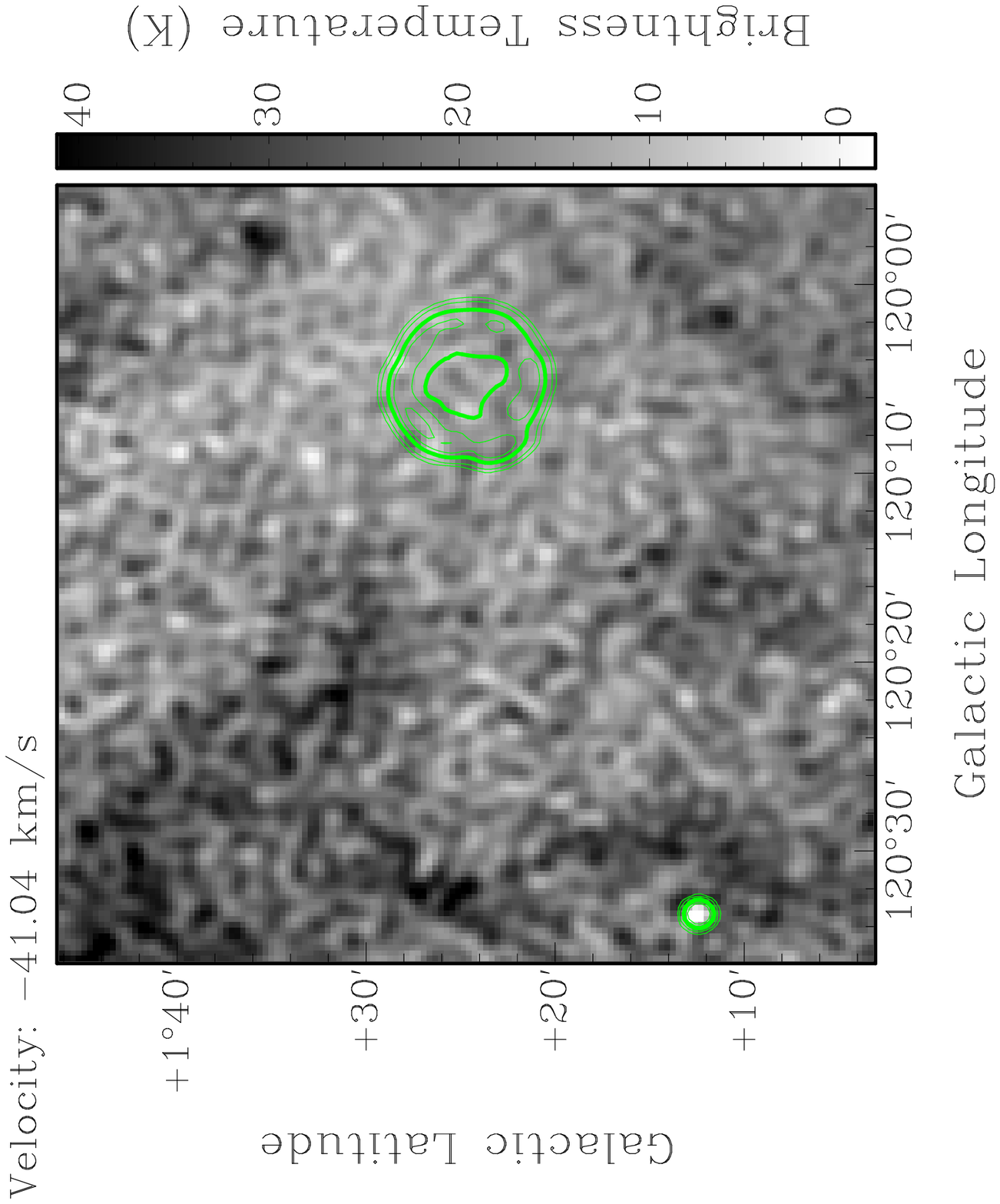}}
\put(210,630){\includegraphics{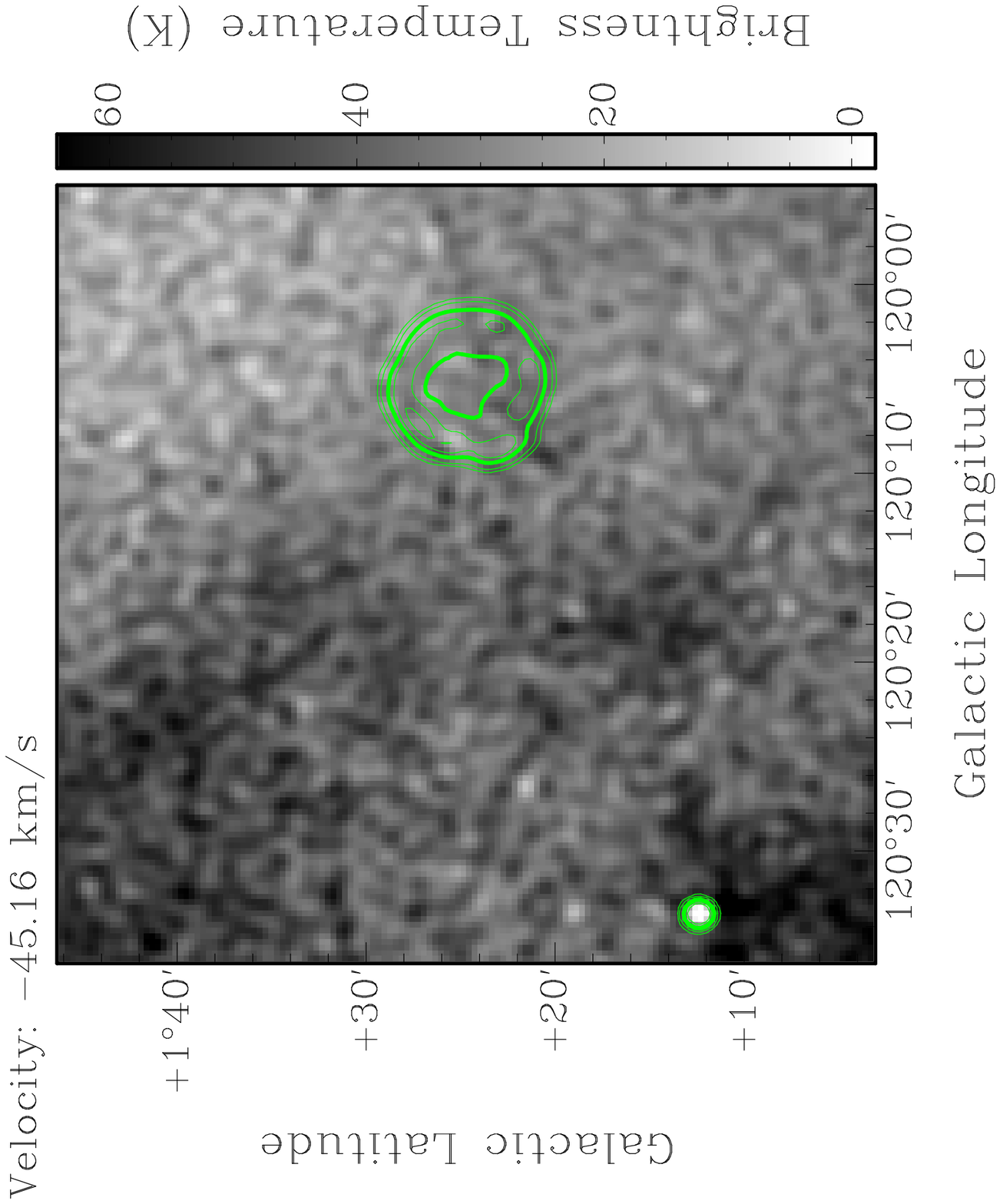}}
\put(-40,475){\includegraphics{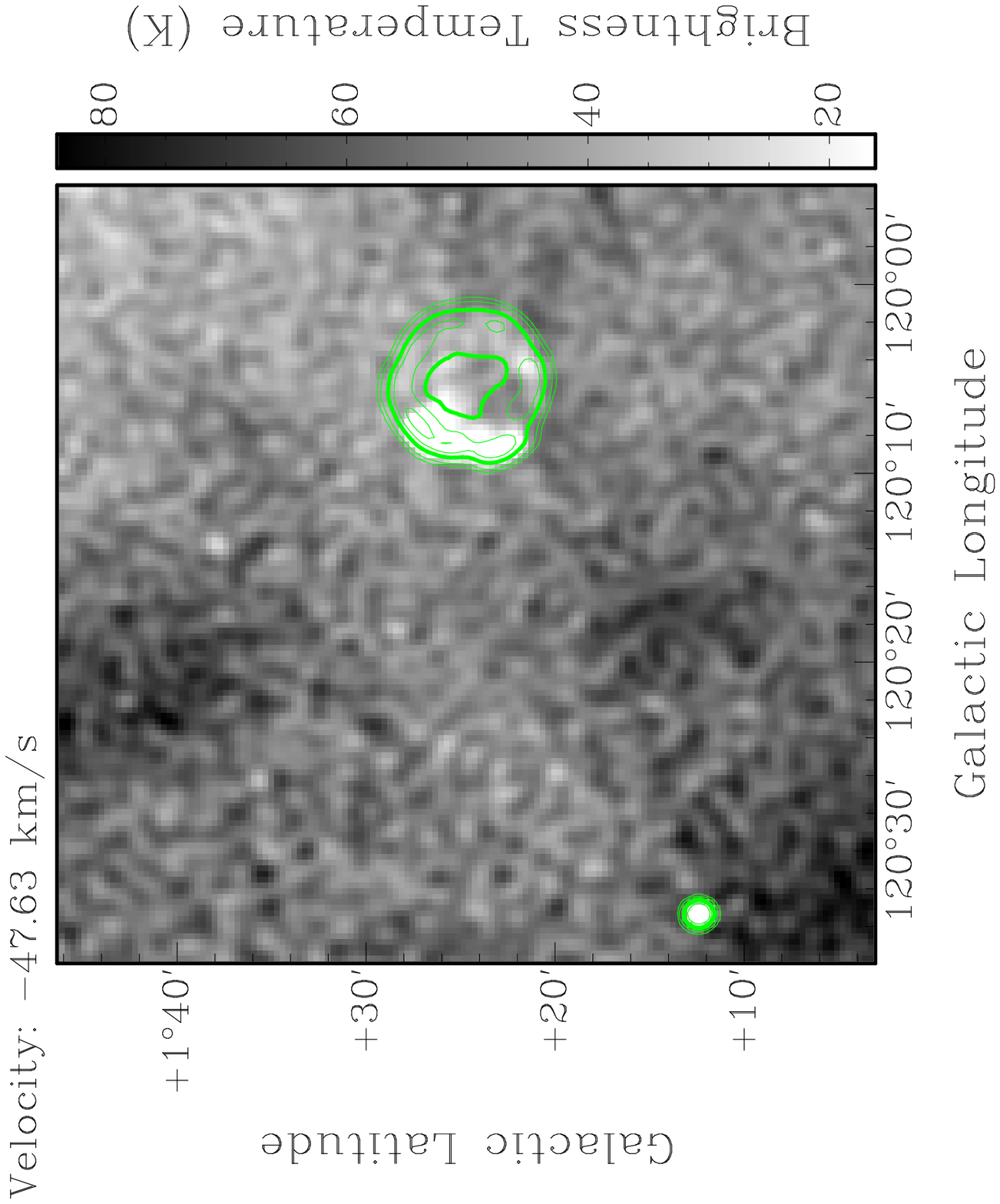}}
\put(210,475){\includegraphics{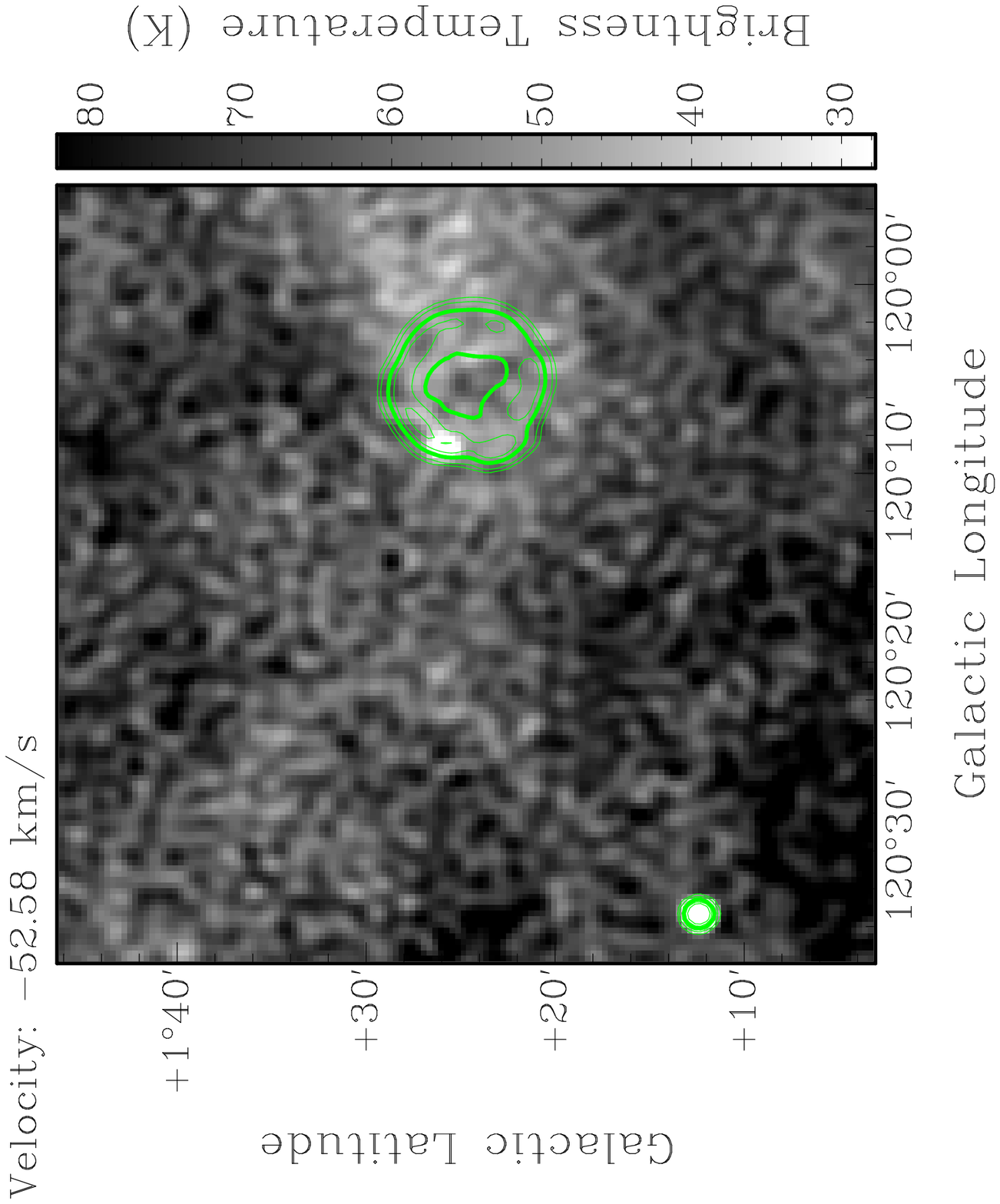}}
\put(-40,320){\includegraphics{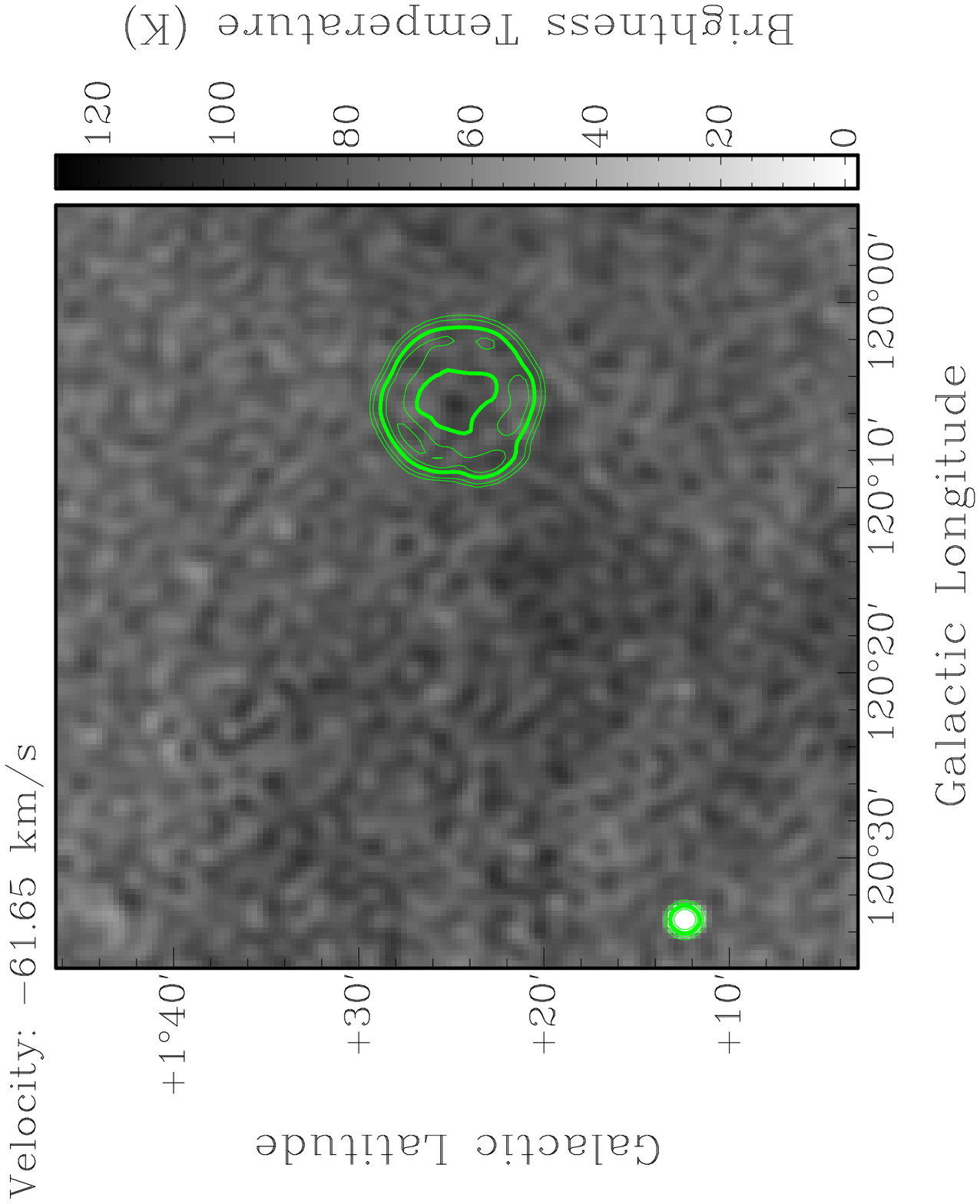}}
\put(210,320){\includegraphics{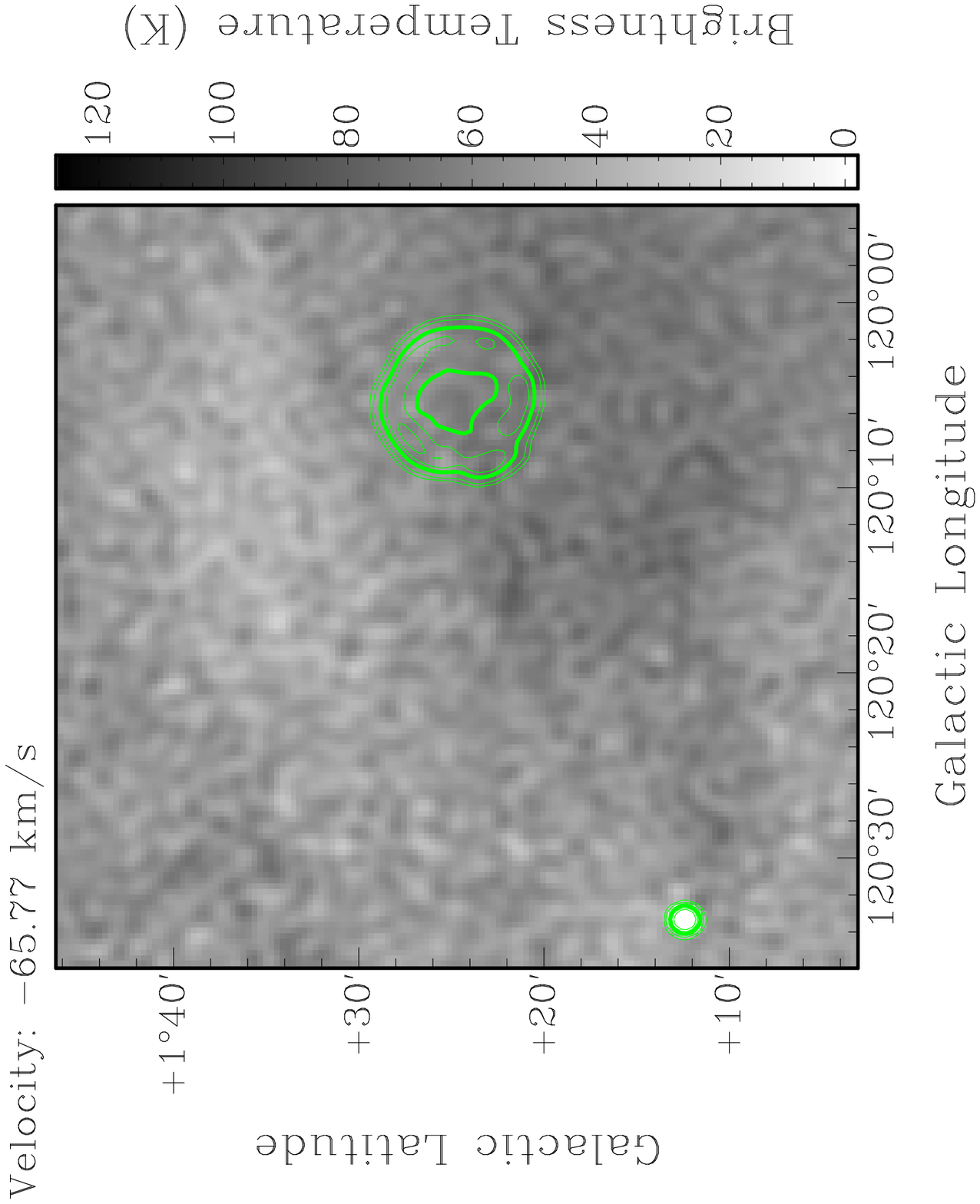}}
\put(-40,165){\includegraphics{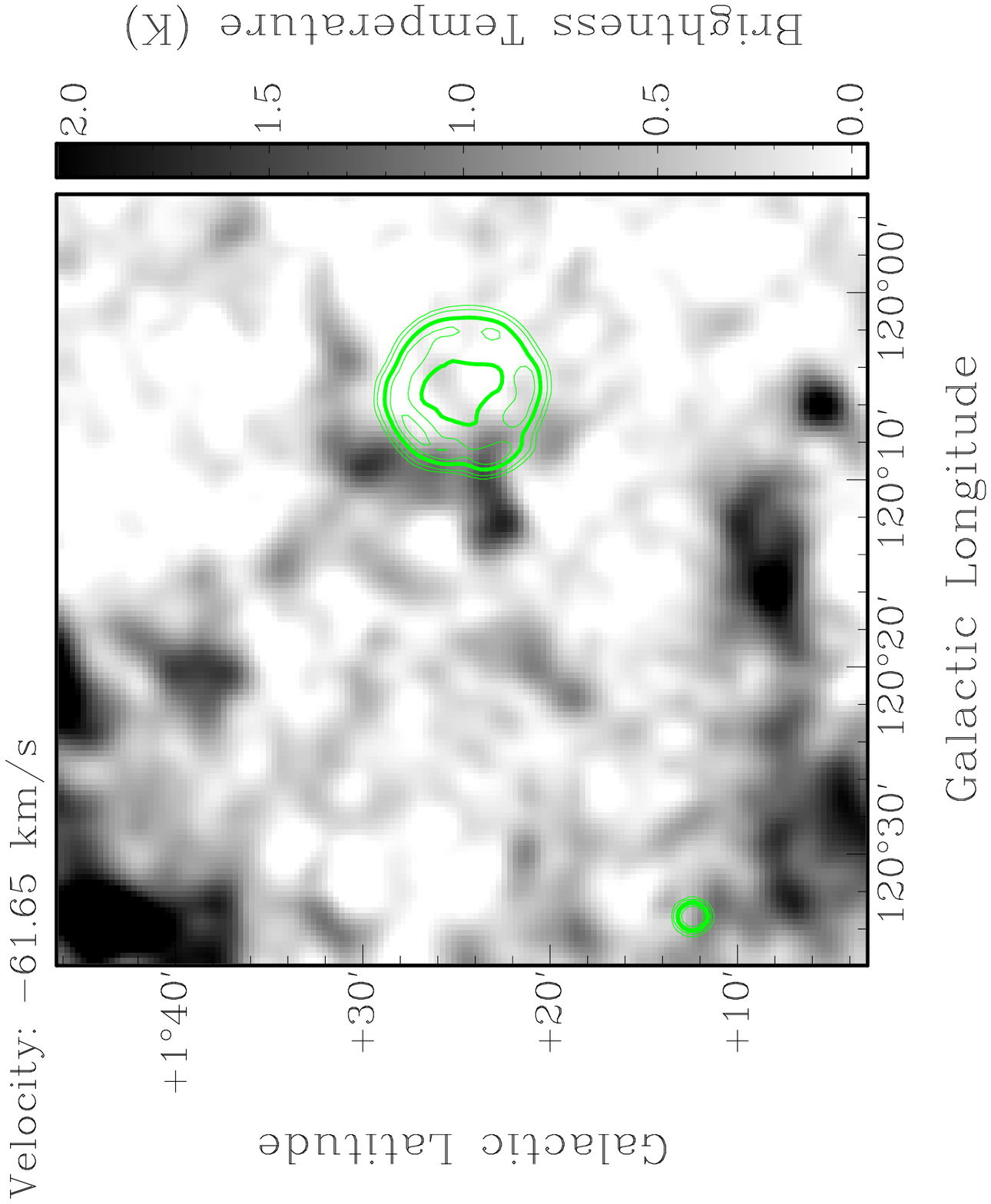}}
\put(210,165){\includegraphics{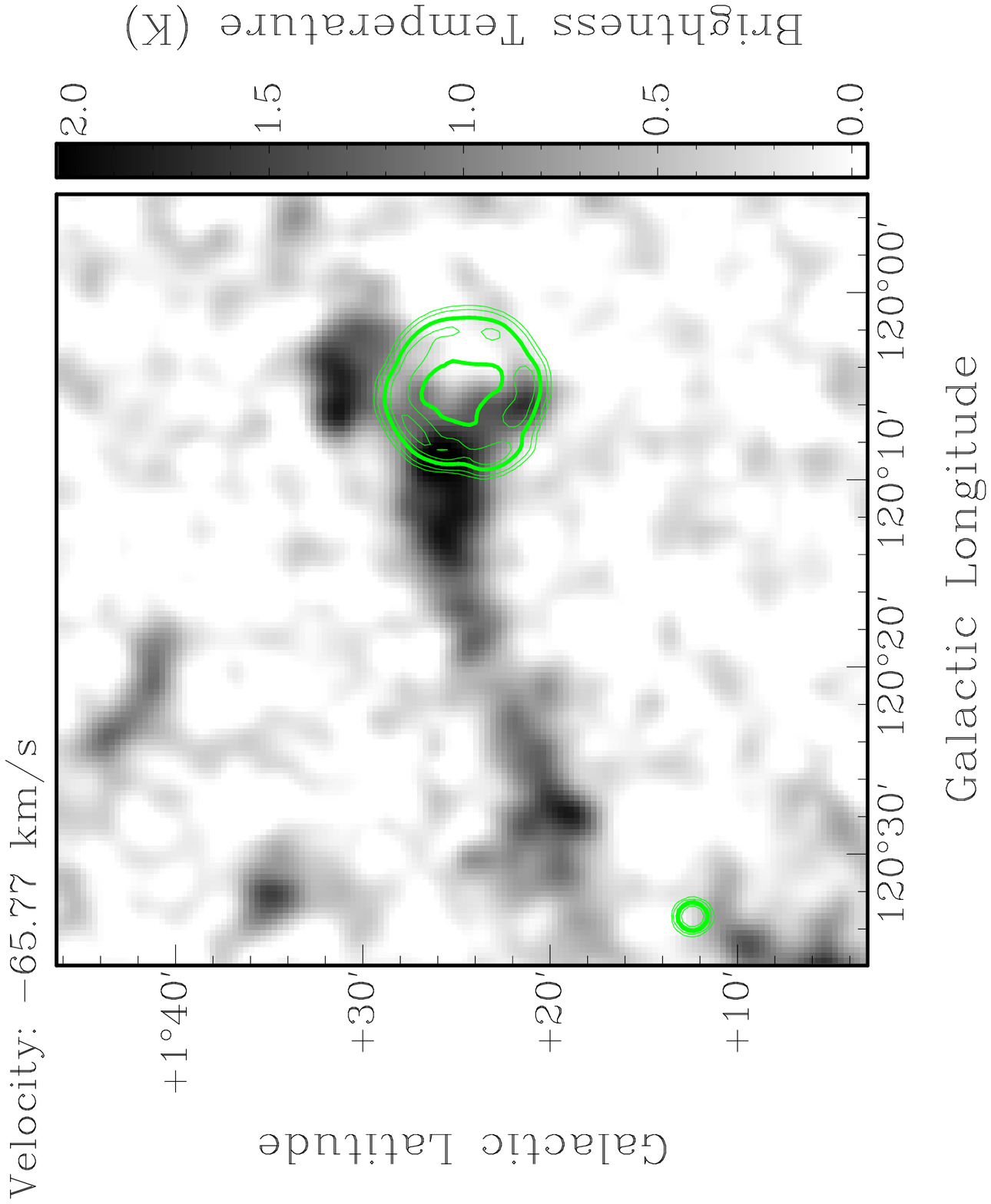}}
\end{picture}
\caption{The first three rows show the HI channel maps. 
The last row shows the $^{12}$CO channel maps. 
 Overlaid in all panels is the 1420 MHz continuum emission with contours: 40, 100, 160, 225 K. The HI map at -47.63 km s$^{-1}$ shows an extended HI along the eastern part of SN 1572. The HI map at -52.58 km s$^{-1}$ shows a small HI clump at the NE site of SN 1572 (see text for detail).} 
\end{figure}

The above results are supported by the HI and CO emission channel maps. We examine all HI and CO channel maps and find that HI absorption towards SN 1572 appears in three velocity windows of: 5 to -20 km s$^{-1}$, -27 to -33 km s$^{-1}$ and -47 to -53 km s$^{-1}$. HI absorption towards the other three compact sources appears in more velocity windows including the above three velocity windows. As an example, Fig. 4 shows 6 HI channel maps and two CO channel maps. The HI maps at velocities of both -61.65 and -65.77 km s$^{-1}$ show clear HI absorption towards G120.56+1.21 but none towards SN 1572. The CO maps show that an extended bright CO cloud (at b = 1.4$^{o}$) appears at all velocities from -61 and -67 km s$^{-1}$ (this can be seen in Fig. 4) and sits on the eastern part of SN 1572's shell. We find HI absorption features at velocities between -54 and -68 km s$^{-1}$ towards G120.56+1.21 but not in SN 1572. This supports the above spectral analysis and confirms that the CO cloud with peak at -64 km s$^{-1}$ is behind SN 1572. 
We also see HI absorption towards SN 1572 in channel maps with velocities in the range of -47.63 and -52.58 km s$^{-1}$. This confirms that HI atomic gas at velocities of -47 to -53 km s$^{-1}$ is in front of SN 1572. 

\section{Discussion and Conclusions} 
\subsection{The CO cloud at -64 km s$^{-1}$ is behind SN 1572}  
We have found that the HI gas at velocities of -47 to -53 km s$^{-1}$ is in front of SN 1572. Both the CO at -64 km s$^{-1}$ and the HI at -60 km s$^{-1}$ do not produce associated HI absorption, so both are likely behind SN 1572. Is it possible that the CO cloud could still be in front of SN 1572 but the amount of cold HI gas in the molecular cloud is too small to produce detectable HI absorption against SN 1572?  

Fig. 2a  shows that the CO at -64 km s$^{-1}$ has associated HI emission (T$_{B}$$\sim$ 75 K). 
Technically, a minimum optical depth of $\tau$ $\sim$ 0.1 may be detected. This requires that HI atomic gas against background continuum source has at least column density of N$_{HI}$=2.3$\times$10$^{19}$ cm$^{-2}$ (N$_{HI}$=1.823$\times$10$^{18}$ $T_{s}$$\tau$$\Delta{v}$; \citet*{Dic90}), given $T_{s}$$\sim$ 25 K (\citep{Sch95} and the FWHM $\Delta{v}$ of the typical HI absorption line $\sim$ 5 km/s. Taking the theoretical mean HI/H$_{2}$ ratio of 0.2 \citep*{Gol05, And09}, this respective molecular cloud has a H$_2$ column density of $\sim$ 1.2$\times 10^{20}$ cm$^{-2}$.

The $^{12}$CO spectrum from box 1 (Fig. 2a) gives W$_{^{12}CO}$ $\sim$ 7 K km s$^{-1}$ for the cloud component. Taking the $^{12}$CO to H$_{2}$ conversion factor of $X$=3$\times$10$^{20}$ [cm$^{-2}$/(K km s$^{-1}$)], this gives a H$_{2}$ column density of $\sim$ 2.1$\times$10$^{21}$ cm$^{-2}$ for the respective cloud. This is enough to produce measurable HI absorption at 18 $\sigma$ level if the cloud component is in front of SN 1572. So the CO molecular cloud at -64 km s$^{-1}$ is behind SN 1572.

\subsection{Does there exist SN 1572-CO cloud interaction?}
 
Previous multi-band observations of SN 1572 \citep{Str82, Gha00, Hug00, Dou01, Hwa02, Lee07, Yan09} have revealed limb-brightened radio and X-ray shell, H$\alpha$ filament along the NE boundary of the shell, the mid-IR emission from SN 1572, suggesting that SN 1572 is surrounded by an inhomogeneous environment. A possible SNR-cloud interaction along SN 1572's NE boundary is able to trigger some observed phenomena, e.g. the NE bright part in radio and X-ray images, the decelerated expansion of the NE rim in radio, optical and X-rays,  the dust emission at the NE boundary, etc. We have concluded that the CO cloud at -64 km s$^{-1}$ is behind SN 1572, but is it possible that the CO is adjacent and interacting with SN 1572?  

This cloud has a H$_{2}$ column density of 2.1 $\times$10$^{21}$ cm$^{-2}$, so its density is $\sim$ 200 cm$^{-3}$, assuming half of the SNR is surrounded by the cloud (R$_{SN 1572}\sim$ 4 arcmin) and taking a distance of 3 kpc (see section 3.5 for detail). This is inconsistent with recent Chandra X-ray observations of the cloud surrounding the remnant. \citet{Cas07} and \citet{Kat10} analyzed X-ray spectra from the thin rim between the blast wave and contact discontinuity, and found that there is little thermal emission from the preshock ambient medium. This requires that the ambient medium density in the vicinity of SN 1572 is less than 0.2 cm$^{-3}$. This is three orders of magnitude lower than the density of the CO cloud. Therefore the CO cloud is not adjacent to SN 1572. We conclude that there is no physical association between SN 1572 and the $^{12}$CO cloud at -64 km s$^{-1}$,  although they are overlapping along the light-of-sight.

\citet{Ish10} detected cold dust IR emission outside the NE and Northwest (NW) boundaries of SN 1572's shell, and suggested that the NE dust emission comes from a possible molecular cloud interacting with the shock front. The origin of the NW dust emission is rather unclear because of the absence of any interstellar cloud nearby. However, our study reveals that the origin of the NE dust emission is also unclear, probably from the molecular cloud at -64 km s$^{-1}$ but unrelated to SN 1572.  We notice that weak TeV emission from SN 1572 is also detected \citep{Acc10}. Although TeV emission from several SNRs has been suggested to originate from interaction between the SNR shock and an adjacent cloud, our result reveals it is not this case for SN 1572 at least.

\subsection{Does there exist SN 1572-HI cloud interaction?}
\citet{Rey99} studied 21 cm spectra in the velocity range of -41 to -106 km s$^{-1}$ towards SN 1572 using Very Large Array (VLA) archive data and single-dish HI observations. They detected HI absorption from -46.4 to -56.8 km s$^{-1}$ towards SN 1572, and found an extended HI absorption along the eastern side of the shell between the velocities of -47.7 and -50.3 km s$^{-1}$. They also found a small high-density HI clump (160-325 cm$^{-3}$) observed as an absorption feature at -51.5 km s$^{-1}$ towards the eastern part of the shell. 
Our study reproduces some of these results: The extended HI structure and the small HI clump seen in our Fig. 4 (at velocities of -47.63 and -52.58 km s$^{-1}$) are found.  
By examining the HI channel maps from -50.3 to -60.6 km s$^{-1}$ shown in their Fig. 2, we see a clear deficit of HI brightness between -52.9 and -56.8 km s$^{-1}$ surrounding SN 1572. This is inconsistent with being a HI absorption feature because it does not correspond with bright continuum emission. It could be either HI self absorption (HISA) or artifacts. 
Our HI absorption spectra clearly reveal reliable HI absorption features in the velocity range of -47 to -53 km s$^{-1}$ but not beyond -53 km s$^{-1}$. Our methods to build HI absorption spectra have reduced false absorption features as much as possible. We do not find any reliable absorption towards SN 1572 in our HI channel maps in the range of -54 to -66 km s$^{-1}$. So any HI absorption feature beyond -53 km s$^{-1}$ is likely not real.

This extended HI cloud along the eastern side of the shell is in front of SN 1572, but is it possible to be adjacent to SN 1572? The HI cloud's density of $\sim$10 cm$^{-3}$ ($\tau$$\sim$0.8, $\Delta$$v$$\sim$3 km s$^{-1}$ from Fig. 2a and 3, T$_{s}$=25K from \citet{Sch95}) is much higher that the ambient medium density of 0.2 cm$^{-3}$ from the X-ray measurements, so the HI cloud is not adjacent to SN 1572.
 
\citet{Rey99} suggested an interaction between SN 1572 and the small NE HI clump because of two factors: the HI clump is near the site of the lowest expansion velocity along the whole shell of SN 1572 (which can roughly explain the slowing of expansion rate of the eastern rim), and a detected H$\alpha$ knot lies near the HI clump. However, \citet{Lee07}, using the Subaru Telescope, obtained a systemic velocity of -30.3$\pm$0.2 to the H$\alpha$ knot G. So we need further new independent observations to distinguish if the knot has relation with the HI clump or not. 

In summary, there is no direct evidence that the extended HI cloud along the eastern part of SN 1572 is physically associated with SN 1572. 

\subsection{Tycho SN 1572, a naked Ia SNR} 
Massive stars have a lifetime of about 10$^{6}$ yr, so Type II/Ibc SNe are expected to take place in a dense, star-forming region where their progenitors are formed. This is different for Type Ia SNe because of the longer time needed for the system to evolve ($\sim$10$^{8}$ yr). SN 1572's  progenitor has wandered $\sim$1 kpc far away from the star-forming site, given an average birth velocity of 10 km s$^{-1}$, therefore is far outside of its parent molecular cloud (generally the size of individual giant molecular cloud is $\sim$ 100 parsec). 
Although SN 1572 might encounter other clouds as it wanders through the interstellar medium, X-ray observations of 1572 show low ISM density. So we believe SN 1572 is likely a naked SNR.

\subsection{Distance to Tycho SN 1572}
The distance to SN 1572 has previously been suggested between 2 to 5 kpc by radio, optical, X-ray and $\gamma$-ray observations (Fig. 6 of \citet{Hay10} shows a summary). As the  Perseus arm is influenced by the spiral shock (leading to a velocity reversal; \citet{Rob72}), it is challenging to estimate kinematic distances to objects in the Perseus arm of the outer Galaxy. This velocity reversal causes a distance ambiguity for gas and objects with radial velocities of -40 km s$^{-1}$ to -55 km s$^{-1}$ in the line-of-sight to SN 1572, for $v \le$ -55 km s$^{-1}$ the radial velocity decreases monotonically with distance (see Fig. 2 of \citet{Alb86}). Previously, HI absorption observations have been made towards SN 1572. \citet{Alb86} made aperture synthesis observations of HI using the Cambridge Half-Mile Telescope, and suggested a distance in the range 1.7 - 3.7 kpc. \citet{Sch95} used the VLA to study HI absorption towards SN 1572 and nearby compact sources including G120.56+1.21 and G119.74+2.4, and estimated a distance of 4.6$\pm$0.5 kpc. The distance difference between \citet{Alb86} and \citet{Sch95} is caused by how to deal with a possible weak absorption feature at -60 km s$^{-1}$. \citet{Alb86} thought the absorption could be either from HI in a turbulent state around a filament or could be spurious caused by small-scale variation in HI emission. \citet{Sch95} believed it is real and SN 1572 is farther than the region of distance ambiguity. 

We obtain HI absorption features with higher quality than previous studies, and find clearly that the highest absorption velocity is -53 km s$^{-1}$ towards SN 1572. Therefore we exclude the large distance of 4.6 kpc. Due to absence of HI absorption between -40 and -45 km s$^{-1}$ towards SN 1572, \citet{Alb86} further proposed that the HI emission between -40 and -45 km s$^{-1}$ could be behind SN 1572, and that SN 1572 most likely is located at the near distance of the major absorption feature at -50 km s$^{-1}$. Anyway, they kept an option that SN 1572 has small probability to be at the far side distance of 3.7 kpc of the same absorption feature.     
 
Figure 2 show that the HI gas at -41 to -46 km s$^{-1}$ has no associated obvious HI absorption towards SN 1572 and definitely produces associated HI absorption towards G120.56+1.21. We notice that the HI channel maps in this velocity range show more HI gas with brightness temperature above 20 K surrounding G120.56+1.21 than SN 1572. Could absence of HI absorption in the velocity range towards SN 1572 be due to insufficient HI gas to produce measurable optical depth in front of SN 1572? However, the HI gas has column density of $\sim$ 3$\times$10$^{20}$ cm$^{-2}$ (N$_{HI}$=1.832$\times$10$^{18}$$T_{B}$$\Delta{v}$, $T_{B}$=35 K, see Fig. 2a) which is enough to produce detectable HI absorption, because $\tau$$\ge$ 0.1 requires a minimum HI column density of $\sim$ 7$\times$10$^{19}$ cm$^{-2}$ (here T$_{s}$=75 K, which is an average value in the low velocity range from \citet{Sch95}. They also noticed that it goes below 25 K near -50 km s$^{-1}$). So this gives an upper limit distance for SN 1572 which is in front of the HI at -41 to -46 km s$^{-1}$. In addition, \citet{Sch95} detected HISA at -49 km s$^{-1}$. HISA is generally produced by foreground cold HI in front of background warm HI at same velocity. Because of the velocity reversal in the Perseus Arm, HISA features are widely observed in the CGPS \citep{Gib05, Tia10}. The HISA at -49 km s$^{-1}$ likely originates from the same cause, i.e. cold HI at near side of the velocity reversal absorbs emission from warm HI at the far side at same velocity of -49 km s$^{-1}$. Because the - 49 km s$^{-1}$ HI is in front of SN 1572 (Figs. 2 and 3), SN 1572 must be between the HI at at near side with -48 km s$^{-1}$ and the HI at far side with -41 to -46 km s$^{-1}$. In other words, the distance to SN 1572 is between 2.5 to 3.0 kpc.  We use the \citet{Fos06}'s model which is similar with the Robert's model but puts the spiral shock front at 2.5 kpc in the direction to SN 1572 (also see Fig. 14 of \citet{Sch95}) 

A kinematic distance of 2.5 to 3.0 kpc is roughly consistent with new estimates from other independent methods. \citet{Vol08} suggested that SN 1572's distance is greater than 3.3 kpc by modeling the existing $\gamma$-ray measurements from SN 1572. \citet{Kra08} estimated distance of 3.8$^{1.5}_{-0.9}$ kpc using classic brightness-distance relation and accounting for interstellar foreground extinction. \citet{Hay10} made new $Suzaku$ observations of SN 1572 and estimated an average spherical expansion velocity of $\sim$4700 km s$^{-1}$. They gave a direct distance estimate of 4$\pm$1 kpc by combining the observed ejecta velocities with the ejecta proper-motion measurement by $Chandra$.

\begin{acknowledgements}
The CGPS is a DRAO product. The DRAO is operated as a national facility by the National Research Council of Canada.
\end{acknowledgements}

\end{document}